# Business Model Canvas Should Pay More Attention to the Software Startup Team

Kai-Kristian Kemell*, Atte Elonen*, Mari Suoranta[†], Anh Nguyen-Duc[‡], Juan Garbajosa[§], Rafael Chanin[¶],
Jorge Melegati[‖], Usman Rafiq[‖], Abdullah Aldaeej**, Nana Assyne*, Afonso Sales[¶],
Sami Hyrynsalmi[††], Juhani Risku*, Henry Edison[‡‡], Pekka Abrahamsson*

*Faculty of Information Technology, University of Jyvaskyla, Jyvaskyla, Finland,
0000-0002-0225-4560, 0000-0002-0587-4431, 0000-0003-0469-6642, 0000-0002-4360-2226
[†]School of Business and Economics, University of
Jyväskylä, Jyväskylä, Finland, 0000-0002-3849-4902
[‡]Department of Business and IT, University of Southeast Norway, Bø, Norway, 0000-0002-7063-9200
[§]Universidad Politécnica de Madrid, Madrid, Spain, 0000-0003-0161-3485
[¶]School of Technology, PUCRS, Porto Alegre, Brazil
0000-0002-6293-7419, 0000-0001-6962-3706
[‖]Faculty of Computer Science, Free University of Bozen-Bolzano, Bolzano, Italy,
0000-0003-1303-4173, 0000-0003-3198-851X
**Department of Information Systems, University of Maryland Baltimore Country, Baltimore, United
States
0000-0002-6405-1750
[††]Department of Software Engineering, LUT University, Lahti, Finland, 0000-0002-5073-3750
[‡‡] Lero, NUI Galway, Galway, Ireland, 0000-0002-9494-8059

*Abstract*— **Business Model Canvas (BMC) is a tool widely used to describe startup business models. Despite the various business aspects described, BMC pays a little emphasis on team-related factors. The importance of team-related factors in software development has been acknowledged widely in literature. While not as extensively studied, the importance of teams in software startups is also known in both literature and among practitioners. In this paper, we propose potential changes to BMC to have the tool better reflect the importance of the team, especially in a software startup environment. Based on a literature review, we identify various components related to the team, which we then further support with empirical data. We do so by means of a qualitative case study of five startups.**

## I. Introduction

Business Models play a pivotal role in early-stage startups, guiding the co-development of software products and customers. The Business Model Canvas (BMC) [14] is a prevalent visual modelling tool used to capture the business model of an organization. It consists of nine non-overlapping elements of knowledge which represent how business is done [10]. It allows a company to model its current or desired future business models. By using BMC, an organization can identify necessary resources and activities for capturing busines value.

Although widely studied in information systems research [12], BMC has not been explored in Software Engineering (SE). Moreover, despite its popularity out on the field, research to empirically analyze BMC's effectiveness or suitability in software startup contexts is scarce. An exception is a study by Ghezzi, which presented Lean Startup Approaches in digital startups [6]. Among these approaches, BMC was the most widely used and practitioners recognized their value in outlining important aspects of business ideas.

The team is considered a key aspect of any startup. Existing startup literature has highlighted various anti-patterns and risks that revolve around the team. Lacking capabilities, for example, is a notable problem for any early-stage startup [17]. However, in BMC [14], the team is not a point of focus, being considered simply one of the many resources required to deliver the value proposition. It is of course reasonable to argue that human resources are prominent in certain business models. However, for startups, human capital is the most important resource.

To provide a starting point for including the team more strongly in BMC and other related tools, we study software startup teams in this paper. We conduct a qualitative multiple case study of software startups, focusing on the teams. Data from the cases are collected with semi-structured interviews. The research question of the paper is formulated as follows:

**RQ:** How can the team perspective be incorporated into Business Model Canvas?

## II. Theoretical Background

In this section, first, we discuss the BMC. Then, we discuss the importance of teams in software startups and in software development in the second subsection.

### A. Business Model Canvas

Osterwalder et al. [14] define business model as something that "describes the rationale of how an organization creates, delivers and captures value". Business models are a commonly used in both business and management [11]. To clarify the concept, Osterwalder et al. developed an ontology to describe business models [15]. To achieve this goal, the authors reviewed the literature and identified nine building blocks: value proposition, target customer, distribution channel, relationship, value configuration, core competency, partner network, cost structure, and revenue model.

Based on the ontology, Osterwalder et al. [14] developed the BMC as a tool to communicate business models. It is commonly used to analyse, describe and design business models. The idea behind BMC was to create a shared language that would allow organizations and entrepreneurs to describe and adjust business models to create strategic options.

### B. Team: The Potential Gap

Software development is traditionally carried out in teams of varying hierarchy and levels of authority. In this context, teams seldom communicate with each other, which may lead to inconsistency on a project level [19]. Agile teams differ significantly from traditional software development teams [7]. They are often seen as self-organizing, and in general, are at the core of software development [7].

In startups, agile methods are often used in conjunction with the Lean Startup method to increase likelihood of success [1]. Though there is no reason to assume that aforementioned success factors such as learning and communication are less important in software startups than other types of software organizations, a key issue typically highlighted in terms of teams in software startups is the lack of capabilities [18].

Based on existing literature, we can identify components related to the team that are considered to affect success in startups. Muñoz-Bullon et al. [13] link startup success with the team, noting that since startups seldom have financial resources to leverage, most of their resources are bound to the team and its capabilities. Especially in early-stage software startups, having the right capabilities in the initial team is pivotal [18]. This would be the first component: resources.

In relation to team-related resources, Muñoz-Bullon et al. [13] also discuss heterogeneity. They note that teams with more heterogeneous capabilities were more likely to generate positive outcomes. Closely related to resources, the importance of personal networks has been highlighted in the context of startups as well [3, 11].

We consider Way-of-Working to be the third component. Aside from specific methods or practices in the context of SE, this refers to working culture. For example, Giardino et al. [5] underline the importance of reactiveness in adjusting to changing market situations or the emergence of new technologies while searching for a business model. In terms of methods, on the other hand, software startups work largely either ad hoc or using differing agile methods and practices [16], but their effect on success has not been verified.

Finally, Karhatsu et al. [9] linked team success with self-guidance and team autonomy, i.e. self-organization. Teams that exhibited high degree of self-organization were more likely to achieve positive outcomes [9].

Thus, we identify at least four possible components related to the team: (1) networks, (2) human resources, or capabilities, (3) way of working, (4) self-organizing. In the empirical portion of the study, discussed next, we seek to validate this list of components, and to see whether any additional components arise from the data.

## III. Study Design

Data for this study were collected from the Startup Laboratory of University of Jyväskylä. The Startup Laboratory is a research unit that teaches and studies software startups, but also provides some incubator services for early-stage software startups. Specifically, the laboratory organizes course-based early-stage software startup incubation.

Five startups were studied. One of the startups was ultimately educational only, i.e. the team came to consider the idea unviable during the incubation. One of the startups went on to become a business for some time but was discontinued later. Three of the startups still exist in some form.

We collected data using two sources. First, we conducted semi-structured interviews with some of the team members of some of the startup. Specifically, we interviewed 5 startup teams and in total 8 respondents. All interviews were recorded, the records transcribed, and the analysis done using the transcripts. The interviews were done either Face-to-Face or via Skype, in English.

The interview questions were split into five categories. First, the respondents were asked about their professional background, and their background in terms of entrepreneurship or startup entrepreneurship in general. Secondly, they were asked questions related to their startup and their role in it, such as its business idea, when it was founded, and whether there had been any pivots. Thirdly, the respondents were asked questions focusing on the team, such as how many members the team had, what their roles were, and why they were included into the team. Fourthly, the respondents were asked questions related to the business model of the startup, as well as areas of the business model canvas such as key partners and unique value proposition. Finally, the respondents were asked questions related to the success of the startup, where success refers to receiving external funding and their plans to continue with the startup.

In terms of analysis, emphasis was placed on differences and similarities of the startups in the form of a cross case analysis. A thematic analysis inspired approach was first utilized to make sense of the data in a systematic manner.

## IV. Results

This section is split into subsections according to the components described in section II.C. In presenting our results, we highlight summarizing findings in the form of Primary Empirical Conclusions (PECs) which we will then discuss further in the following section. The interview citations are presented as-is, as spoken by non-native English speakers. Moreover, it should be noted that the analysis is not based on these individual citations alone but a cross comparison of the cases.

### A. Component 1: Network

On multiple occasions, the respondents discussed the importance of personal networks. Given that especially early-stage startups operate with scarce financial resources, having personal networks to leverage for various purposes was considered highly important. Even though operating within the Startup Laboratory gave the teams some contacts to leverage by association, all but one team also highlighted the importance of their own, external networks.

*"Then of course at the top of the cake we have team member X, and team member X is our treasure. He has provided us so much contacts and he has boosted us into what we are doing. And I think he is one of the reasons why we can be successful."* (Interview)

Networks were used by the startups of the respondents to find required human resources, ranging from new team members to advisors or mentors. While the perceived importance of networking varied by team, a recurring theme in the responses of the respondents was a lack of capabilities within the team. A perceived lack of capabilities within the team was largely directly associated with limited personal networks by the respondents themselves.

While we did not gather any data that could be used to objectively validate success factors in this context, the teams themselves considered networking to have been a success factor for them. Vice versa, some teams considered their lack of extensive personal networks to be a shortcoming that affected their chances to succeed negatively.

**PEC1:** Professional networks play a vital role for gathering necessary technical competence for software startups in their early stages.

### B. Component 2: Resources

In discussing resources, four out of five of the startup teams cited the team as their key resource. The respondents felt that their team was their primary selling point as opposed to their product itself.

*"I would say that [another team member] is an experienced entrepreneur, most startups tend to be inexperienced so having a person with actual experience does make the startup somehow unique. From the point of investors as well."* (Interview)

In relation to the network theme discussed in the previous subsection, human resources were also considered key resources in their absence. Lacking capabilities were an issue discussed by many of the respondents.

Lacking time due to other responsibilities was another recurring topic in the interviews. As many of the respondents were students, part-time or full-time, they were balancing their startup activities with other responsibilities. This was especially true for those that had no revenue or investors. In those cases, the team members had to either study or work.

To summarize, the team was considered the most important resource the startups had. Even if the ideas were what they wanted to sell, they felt that it was important that they were the right people to carry out that idea.

**PEC2:** The capabilities of the startup team itself are the most crucial resources for an early-stage startup. The necessary team capabilities of an early stage startups include business development knowledge, previous business experience, and key technical resources.

This observation is related to the PEC1 that networks are a success factor for early-stage startups. However, not all team members have to have extensive personal networks to leverage, even if it would hardly be a negative situation either, arguably. Rather, these two PECs combined result in a further observation, which is grounded in extant literature, as we discuss in the next section:

**PEC3:** A heterogeneous team, capability-wise, is a success factor for early-stage startups.

*C. Component 3: Way of Working*

While all the startups had established some agreed-upon way of working, they scarcely followed any existing methods for working. The ways of working in the startups were considered an iterative process that varied between tasks and was adjusted as needed, rather than a pre-planned method.

In some of the startups, the team members did not have clear roles assigned to them at all. This was the case in the startups that lacked the most resources in terms of capabilities. As they did not have enough team members specializing in different tasks, they simply worked on whatever tasks they had that required work. In these cases, the roles of the team members changed based on what was required at the time.

*"I don't know if we had any specific roles. We worked well as a team and just managed the tasks we had, prototypes or connections or people anything we had. I don't know. Did we have any specific roles? What would you say? [Asking other respondent]"*

*"[Replying] Yeah, we were not like that every time some kind of work came up, it wouldn't be like you are better at doing this and I am better in this, we didn't really think like that..."* (Interview)

The ways of working of the respondents' startups worked ad hoc, adjusting to the situation at hand. The teams did not follow any clear methodologies and seemed to work iteratively, adjusting their way of working as well as their idea based on the context that they were in.

**PEC4:** Early-stage startups work largely ad hoc and the available human and other types of resources defines the team's way of working, rather than the founders' own wills.

*D. Component 4: Self-Organizing*

Given that the startups of the respondents consisted of small teams of 1 to 5 members, decision-making was generally carried out between the entire team. Everyone on the team would have their say on any non-task-specific decision being made, should they have felt like doing so.

*"We have been making those decisions together with the second co-founder"* (Interview)

Self-organization was not always considered a positive factor by the teams. Some startups were not self-organizing by choice, but rather, because they had to be. If they made no decisions themselves, no one made any. For an early-stage startup, arguably, external pressure in decision-making mostly comes into play once, or if, external funding is acquired.

**PEC5:** Early-stage software startups are often forced to be self-organizing, as the team usually only has a few members and no external funding.

V. DISCUSSION

Overall, our findings served to validate the research framework devised based on extant literature in section III. While the importance of the team is to some extent acknowledged in extant literature [2][20], our findings further highlight this. The novelty of our findings stems from the argument that the focus on software startups should be on team first rather than idea first. This is also the case in the context of the Business Model Canvas.

PECs1 and 2 supports the notion that networks are important for software startups present in extant literature, and that the team is the most important resource of a startup [4] [13]. PEC3 reinforces the idea described by Muñoz-Bullon et al. [13] that the more heterogeneous the team is, the more likely it is to create positive outcomes. Our data supports it.

PEC4 confirms the notion of Paternoster et al. [16] who argued that software startups either use various agile methods and practices or simply operate ad hoc. Especially early-stage software startups seem to lean towards working ad hoc. It is likely that as the development team grows and more developers come on board, the team also begins to utilize some tangible methods or practices, though our data does not help us ascertain this assumption.

Self-organization is considered a success factor in agile teams [9]. However, as stated in PEC4, these startup teams ultimately did not work in an agile manner, making our findings not directly comparable to those in existing literature. Indeed, the startup teams in this study were not self-organizing by a conscious choice. Rather, they were self-organizing because they had to be.

The primary limitations of this study stem from the qualitative case study method. As argued by Eisenhardt [3], however, it is a suitable approach for novel areas of research, as well as to gather more in-depth data in general. The multiple case approach further reduces the threat to validity stemming from this approach. Moreover, our findings ultimately served to validate existing research for the most part, and thus are backed by extant studies.

VI. CONCLUSIONS

In this paper, we have studied software startup teams from the point of view of the Business

Model Canvas. We have done so by interviewing eight respondents from five early-stage software startups. The goal of this study was to understand how BMC could be tailored to better suit early-stage software startups, which are one prominent user group of the tool. While existing literature and anecdotal practitioner wisdom highlight the importance of teams in software startups, the tool places little emphasis on the team, which is mostly filed under the 'key resources' category of the tool and never directly mentioned in the description of the category.

Based on our findings, we suggest that the canvas should at least place more emphasis on the team under the existing key resources category. Specifically, the component should emphasize the importance of having a team with the capabilities to carry out the idea of the startup. Alternatively, the team could even be a category of its own in BMC, as it does not receive enough emphasis being a subcategory of an existing category given its importance for software startups.

Regarding future studies, we advocate for more studies on software startup teams. One avenue that the authors themselves intend to pursue is to focus on developing and exploring the role of the team in business models by suggesting how the canvas could be improved to cover this aspect. Afterwards, the new canvas could be evaluated empirically by startups. The team might be the most important resource for a software startup in this knowledge-intensive field. As this study suggests, the team and its capabilities should also be of interest venture capitalists. Therefore, a separate canvas or modelling tool to focus on the team specifically could be of interest for communication between different stakeholders' interest in the startup.

ACKNOWLEDGEMENT

This project has received funding from the European Union's Horizon 2020 research and innovation programme under the Marie Skłodowska-Curie grant agreement No 754489 and with the financial support of the Science Foundation Ireland grant 13/RC/2094.